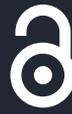
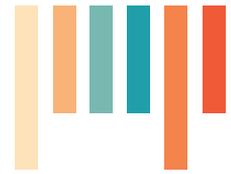

Philosophy of Physics

# The Simply Uninformed Thermodynamics of Erasure

**JOHN D. NORTON**

**RESEARCH**

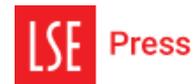

## ABSTRACT

1. Strong and weak notions of erasure are distinguished according to whether the single erasure procedure does or does not leave the environment in the same state independently of the pre-erasure state.
2. Purely thermodynamic considerations show that strong erasure cannot be dissipationless.
3. The main source of entropy creation in erasure processes at molecular scales is the entropy that must be created to suppress thermal fluctuations ("noise").
4. A phase space analysis recovers no minimum entropy cost for weak erasure and a positive minimum entropy cost for strong erasure.
5. An information entropy term has been attributed mistakenly to pre-erasure states in the Gibbs formalism through the neglect of an additive constant in the "–$k$ sum $p$ log $p$" Gibbs entropy formula.

**CORRESPONDING AUTHOR:**

**John D. Norton**

Department of History and Philosophy of Science, University of Pittsburgh, Pittsburgh, PA, USA

jdnorton@pitt.edu

# 1. INTRODUCTION

In 1929, Leo Szilard (1929) imagined a cylinder containing a gas of a single molecule at thermal equilibrium with its environment at temperature *T*. A partition is inserted and divides it into two parts, trapping the molecule on one side. If we conceive of this partitioned cylinder as a memory device recording either "left" *L* or "right" *R*, we can ask for the dissipation, that is, the minimum entropy created in returning it to a reset state, such as *L*, as shown in Figure 1.

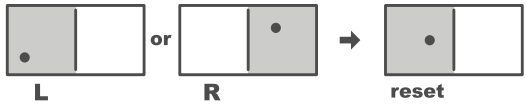

**Figure 1** Erasure of a Szilard One-Molecule Gas System.

This standard example will be used to illustrate more general results about erasure in systems at molecular scales, that is, those in which quantities of heat and energy are of the order Boltzmann's constant *k*.

The entropy cost of erasure has been the locus of a literature that employs notions of information and computation. Szilard located an entropy cost in acquiring the information that identifies the location of the trapped molecule. Landauer (1961) and Bennett (1987) treated the relevant systems as computational devices that process information and located an entropy cost in the many-to-one logic of erasure. This focus on information and computation has, I believe, served only as an unproductive distraction.[1] The many-to-one mappings of erasure can be treated quite adequately without considerations of information and computation, as will be shown in the simpler, uninformed accounts of earlier Sections 2 to 5 below. Later, Sections 6 and 7 will show that the identification of information entropy with thermodynamic entropy is mistaken and that treatments of erasure that depend upon it produce spurious results.

The phase space analysis of Sections 4 and 5 adopts the assumptions of the "Boltzmannian" approach to statistical physics. Most notably, it assumes that the dynamical evolution of a system's phase point is such that the probability of its being in some volume of the total system phase space is proportional to the phase space volume. Sections 6 and 7 draw on the "Gibbsian" approach. It employs the "Gibbs entropy" and the demonstration introduced by Gibbs and Einstein of the connection between Gibbs entropy and Clausius' thermodynamic definition for entropy in terms of heat. There are enduring, unsolved foundational problems in both approaches. Frigg and Werndl (2023) provides a recent, accessible survey of them. The existing literature on erasure draws freely on these approaches and reasonably so, since these approaches have enjoyed considerable success in recovering the generic behavior of thermodynamic systems. The project of this paper is not to address and solve these foundational problems, but to ask what we can learn if these approaches are applied to erasure processes.

# 2. CONDITIONS ON ERASURE

A transformation that takes either of two distinct states, such as *L* or *R*, to a reset state, such as *L*, is not by itself an erasure. The existing literature provides two additional conditions:

---

[1] It has, as shown in Norton (2018), made it easy to overlook a simple and serviceable exorcism of Maxwell's demon that employs no information or computational notions.

> *Szilard's condition.* The erasure must be a single procedure, specified independently of which state is presented for erasure.

This condition was fundamental to Szilard's (1929) attempt to use the gas to create a Maxwell's demon, in which he needed to expand the gas reversibly and isothermally. The obstacle was that different mechanical couplings were needed, according to whether the molecule was trapped on the left or the right side of the partition. *Prima facie*, two different procedures were needed. Szilard sought to collapse them into a single procedure by including in the procedure the detection of the location of the molecule that enabled the appropriate mechanical coupling to be deployed.

> *Bennett's condition.* The completed erasure procedure must leave the environment in the same state, independently of which state was presented for erasure.

If a many-to-one mapping leaves the environment in a different state according to which of the initial states was presented for erasure, then a trace of that original state remains. In computational terms, the data has not been erased but merely relocated. This condition is associated with Charles Bennett for his introduction of the notion of reversible computation. He sought to avoid Landauer's (1961) conclusion of an inevitable dissipation associated with erasure. Bennett's (1973) proposal was that data in one location could be erased if a copy of the data was secreted elsewhere in a reversible process that, in Landauer's analysis, could be effected non-dissipatively. Locally, the data would be erased, but not globally. Reversing the process would recover the erased data from its remote storage.

The importance of taking the global perspective is central to Bennett's (1987) exorcism of the Maxwell demon implicit in Szilard's problem. Bennett argued *erroneously*[2] that a dissipationless measurement of the position of Szilard's molecule was possible and that it would enable a thermodynamically reversible resetting of Szilard's cylinder. It would be a dissipationless erasure. The catch, Bennett argued, was that the device implementing the erasure must record the location of the molecule in its memory. Completing the cycle requires erasing the memory, which, according to Landauer's analysis comes, with a cost of $k \log 2$ of entropy. Overall, the erasure is not dissipationless.

What I shall call *strong erasure* satisfies both conditions. *Weak erasure* satisfies only *Szilard's condition*. The weak notion, if it can be realized, may be useful in more practical applications in which the violation of *Bennett's condition* consists of a very slight difference in the heating of the environment, according to the pre-erasure state.

## 3. NO DISSIPATIONLESS, STRONG ERASURES IN THERMODYNAMICS

The simplest theoretical analysis of erasure arises when we represent the system within a state space with the variables of ordinary thermodynamics: pressure $P$, volume $V$, internal energy $U$, entropy $S$, and so on. The one-molecule gas is represented as a continuous fluid, filling the volume accessible to it, conforming with the ideal gas law, $PV = kT$, where $k$ is Boltzmann's constant. The familiar laws of thermodynamics apply. Within this impoverished representation, no process corresponds to Szilard's insertion of the partition, since that process is indeterministic and contrary to the second law in reducing entropy. In phenomenological thermodynamics, inserting the partition would merely divide the gas fluid in half.

---

2 Bennett's detection device will be fatally disrupted by fluctuations, as noted in Earman and Norton (1999, 13–14).

The result below in thermodynamics precludes a non-dissipative strong erasure. While I believe that something like it has long been implicit in discussions of erasure, I hope that it is useful to spell it out more fully so that its precise content is visible. The result is relevant to a statistical mechanical account of thermal systems. Such an account must either return the thermodynamic result in a suitable limit or give reasons for its failure.

### 3.1 DERIVATION

The two conditions for strong erasure are implemented in a thermodynamic analysis concerning a system "*Sys*," such as Szilard's one-molecule gas, and the environment "*Env*" with which it interacts. A procedure "*P*" includes familiar operations on thermal systems, such as heatings and coolings, compressions and expansions. It will evolve the pair from an initial state 1 to the final state 2. This evolution is written as

$$(Sys_1, Env_1) \rightarrow_P (Sys_2, Env_2).$$

The properties assumed for these processes are:

> *Assumption 1: Determinism*. The unique state $(Sys_2, Env_2)$ to which $(Sys_1, Env_1)$ evolves is fixed by $(Sys_1, Env_1)$ and $P$.

This assumption requires that the specification of the environment be sufficiently expansive as to include all parts that may affect the course of the process. What is precluded are indeterministic or stochastic evolutions in which an initial state may evolve under *P* in uncertain ways to different final states. A common representation of such a deterministic process is a single curve in the thermodynamic state space connecting initial and final states.

> *Assumption 2: Reversibility.*[3] There are special cases of processes for which there are reversed processes that trace out the same time evolution of the system and environment states, but in the reversed order.

In such processes, the thermodynamic entropy of the combined system and environment is constant. Since dissipation here just means creation of entropy, they are the least dissipative processes. If we represent a possible reversible process as

$$(Sys_1, Env_1) \rightarrow_{P,rev} (Sys_2, Env_2),$$

then the assumption assures us of the possibility of a second process:

$$(Sys_2, Env_2) \rightarrow_{P',rev} (Sys_1, Env_1).$$

The reversed procedure *P'* is realized by reversing the direction of heat and work transfers of the original process *P*.

These assumptions support the following results:

> *Result 1. No reversible forks.* We cannot have both of the processes with the same procedure *P*:
> $(Sys_1, Env_1) \rightarrow_{P,rev} (Sys_2, Env_2)$ and $(Sys_1, Env_1) \rightarrow_{P,rev} (Sys_3, Env_3)$.

---

3   Norton (2016) has argued that reversible processes cannot be the evolution of a single state since the assumption of the perfect equilibrium of driving forces precludes change. Rather, talk of a reversible process is an abbreviated reference to a collection of real, dissipative processes such that limit operations return the properties associated the reversible process. The abbreviation will be employed here without further apology, since the complications of the more careful analysis will not alter the outcomes of the analysis.

This follows since the evolution is deterministic and, if a reversible process has taken ($Sys_1$, $Env_1$) to a later state ($Sys_2$, $Env_2$), this same process cannot also take it to a different state ($Sys_3$, $Env_3$). Moreover, this one process cannot take ($Sys_1$, $Env_1$) also to ($Sys_3$, $Env_2$), where we have set $Env_3 = Env_2$. A second result follows if we apply the condition of reversibility to *Result 1*.

> *Result 2. No reversible many-to-one processes (no strong erasure).* We cannot have both of the processes with the same procedure *P*:
> ($Sys_2$, $Env_2$) →$_{P,rev}$ ($Sys_1$, $Env_1$) and ($Sys_3$, $Env_2$) →$_{P,rev}$ ($Sys_1$, $Env_1$).

For if we assume otherwise and if we apply the condition of reversibility to these many-to-one processes, we recover a forked process prohibited in *Result 1*. *Bennett's condition* is applied in requiring that the environmental states are the same after the process is completed.

This result does not preclude many-to-one processes such as erasure. Rather it precludes strong erasure from being implemented by reversible, that is, non-dissipative, processes. The thermodynamic analysis of this section does not preclude implementation of weak erasure by a reversible process.

## 3.2 LIMITATIONS AND EXTENT OF APPLICATION

Purely thermodynamic analysis does not, I believe, have the means to assign a positive lower bound to the amount of entropy that must be created in strong erasure. This limitation is supported by the fact that statistical mechanical results must revert to thermodynamic results in the limit of vanishingly small Boltzmann's constant *k*. Analyses within statistical mechanics derive positive lower bounds on entropy creation that are linear functions of Boltzmann constant *k*, such as the *k* log 2 commonly cited in Szilard's problem. If we assume an arbitrarily small *k*, then these lower bounds to entropy creation become correspondingly small and have no non-zero lower bound.

In spite of these limitations, this thermodynamic result already ensures the failure of proposals for dissipationless strong erasure that only employ procedures that can be realized within phenomenological thermodynamics. This set is expansive and includes reversible heating and cooling, the reversible compression and expansion of the volume degrees of freedom of any thermal system, the reversible manipulation of the electric and magnetic properties of continuous media and a multiplicity of reversible processes applied to such continuous thermal systems undergoing phase transition. No combination of these processes, no matter how ingenious, can effect a dissipationless strong erasure, as long as the system is treated like one within phenomenological thermodynamics. In the case of Szilard's problem, this means that the processes treat the one-molecule gas as a continuous fluid with the equation of state $PV = kT$. No procedure can implement strong erasure by any combination of reversible heatings or coolings or expansions or contractions, isothermal, adiabatic, or otherwise.

## 4. THE PHASE SPACE ANALYSIS

We can accommodate the statistical mechanical character of thermal systems by exploring their properties in a phase space analysis. It will be "Boltzmannian" in character. The totality of the system and its environment is represented by a single point in the phase space; and the evolution in time of the phase point is governed by an unmanipulated[4] Hamiltonian. Hence, the time evolution of the point representing the totality is restricted to a surface of constant energy in the total phase space. Since *Szilard's condition* requires that a single procedure is used for erasure, it follows that the same Hamiltonian must be used, no matter which state is presented for erasure. This requirement plays a central role in the literature in establishing the existence of non-trivial lower bounds on dissipation. It is important in Myrvold's (2021) analysis and again in the quantum dynamical recovery of lower bounds in Anderson (2022, 5–7, 11).

Drawing from the Boltzmannian approach, the Boltzmann entropy $S$ of a state is defined by the phase volume $V_{ph}$ that represents the state as

$$S = k \log V_{ph}. \qquad (1)$$

An independent assumption, distinctive of the Boltzmannian approach, is that the dynamical evolution of the phase point is such that the probability $P$ that the system point will be in any given volume $V_{ph}$ of the phase space is proportional to its volume:

$$P \sim V_{ph}. \qquad (2)$$

The extent to which real systems conform with this assumption remains a topic of extensive debate in the literature on the foundations of statistical physics.

Liouville's theorem of Hamiltonian mechanics asserts that volumes of phase space are preserved under Hamiltonian evolution. This preservation is incompatible with the Boltzmannian assumption that systems evolve to states of higher entropy and thus of greater phase volume, according to (1); and that they do so with greater probability, according to (2).

The standard solution within the Boltzmannian approach is to divide the phase space into coarse-grained volumes. They are then used to identify the state of a system whose phase point lies within a coarse-grained volume. For applications here, I believe it is adequate to identify a coarse-grained volume as the set of phase points compatible with some set of the macroscopic, thermodynamic variables. Consider, for example, a compressed volume of an ideal gas that expands to fill an otherwise evacuated and isolated chamber. It follows from Liouville's theorem that the volume of phase space accessible to the gas' phase point does not increase. Rather the accessible volume is drawn out into massively convoluted tendrils whose spatial degrees of freedom penetrate all parts of the larger chamber. The volume of that chamber is the macroscopic, thermodynamic variable that characterizes the spatial volume degrees of freedom of the gas. In this sense, the gas volume has increased and, when (3) is applied to the phase volume increase, it has done so with very great probability.

The most important characteristic of the Boltzmannian approach is that a process only advances with probabilistic assurances from an initial to a final state if the phase volume of the final state is significantly larger than that of the initial state. This phase space expansion corresponds to an increase in thermodynamic entropy and is the principal source of dissipation for all processes, erasure or otherwise, at molecular scales. We shall see below that it manifests in more familiar terms as the entropy creation needed to suppress the disruptive effects of thermal fluctuations.

---

4   In a Gibbsian analyses, a process might be represented by a Hamiltonian that varies over time as a function of an externally manipulated parameter. This one varied Hamiltonian can represent multiple procedures and thus violate *Szilard's condition*. If it represents the rightward shift of the partition in a Szilard one-molecule gas cylinder, different mechanical couplings are needed according to whether the gas is trapped on the left and is expanding or the gas is trapped on the right and is compressed.

The near universal practice in the present literature is to consider just the dissipation associated specifically with the many-to-one mapping of erasure. It ignores or mistakenly discounts these fluctuations as nuisances that can be idealized away without compromising the analysis. Because the practice is so wide-spread, the following will treat the dissipation specifically arising from the many-to-one mapping of erasure in the present in Section 4; and then treat fluctuations in Section 5.

The idea that changes in phase space volume determine an entropy cost of erasure has appeared often in the literature, but commonly only as a suggestive slogan. A more careful analysis, such as Oriols and Nikolic (2023, especially Figure 4), shows how coarse-graining must be considered if we are to recover the entropy costs of strong erasure. Turgut (2009) gives a similar if more complicated analysis. Hemmo and Shenker (2012, especially Ch. 12) investigated the same processes at some length from the phase space perspective. They do not arrive at a definite entropy cost for strong erasure because of concerns that the coarse-grained macrostate is not uniquely defined.

## 4.1 WEAK ERASURE

Considerations of many-to-one mappings require no dissipation for the case of weak erasure for the simple reason that weak erasure does not require a many-to-one mapping. Consider a system initially in one of two distinct states, such as the *L* and *R* states of a one-molecule gas, and a *reset* state of equal phase volume. In weak erasure, both systems must evolve under the Hamiltonian to the same reset state. However, their environmental degrees of freedom can remain distinct so that the phase volumes associated with each state can remain the same in magnitude. In that case, it follows from (1) that there is no increase in entropy in each of the system and environment individually; and thus no heat is transferred from the system to the environment. The process is illustrated in the phase space of the highly stylized Figure 2. System degrees of freedom are represented horizontally; and environmental degrees of freedom are represented vertically.

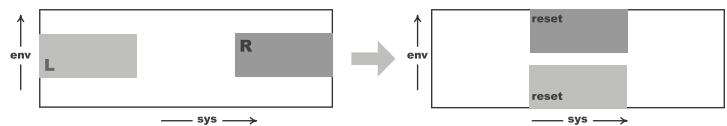

**Figure 2** Weak erasure in phase space.

This weak erasure, by design, does not conform with *Bennett's condition*. If we neglect the dissipation required to suppress fluctuations, we can display a highly idealized, weak erasure procedure for the case of a Szilard one-molecule gas.[5] Assume that the horizontal position only of the molecule in ordinary space in the divided gas cylinder is taken to be the system. Its vertical position is regarded a part of the environmental degrees of freedom. Then a thermodynamically reversible erasure procedure conforming with *Szilard's condition* simply rotates the cylinder by ninety degrees as shown in Figure 3.

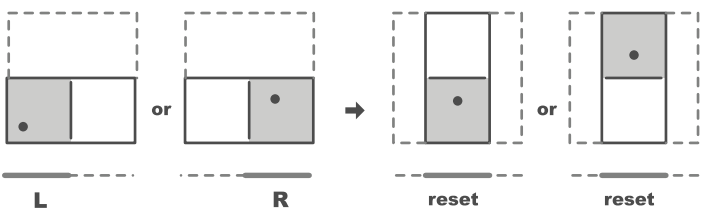

**Figure 3** Weak Erasure of a Szilard One-molecule Gas System.

---

5   I believe this procedure was suggested to me by someone in informal communications and, if could recall who it was, I would credit them.

While this procedure satisfies the formal definition of weak erasure, it does not realize the interesting case of erasure of a Szilard one-molecule gas where the trace of the erased state lies in a slight differential heating of the environment. I know of no procedure, conforming with *Szilard's condition*, that does this.

## 4.2 STRONG ERASURE

A phase space analysis does show an unavoidable entropy cost in strong erasure, which must conform with both *Szilard's* and *Bennett's conditions*. If we take the initial states $L$ and $R$ to be distinct, each state and their associated environments will be represented by disjoint sub-volumes of the phase space $V_{ph,L}$ and $V_{ph,R}$; and the reset state corresponds to another sub-volume $V_{ph,reset}$ that is not necessarily disjoint from the first two states in the system properties.[6] For strong erasure, under Hamiltonian evolution, both system and environmental degrees of freedom must evolve to the same overall *reset* state. We might imagine that it must map the points in the volumes $V_{ph,L}$ and $V_{ph,R}$ to those in the reset state $V_{ph,reset}$, as shown in Figure 4.

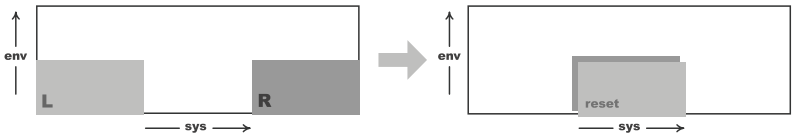

**Figure 4** A Failed Many-to-One Mapping.

If this last specification is correct, it is already enough to assure us that strong erasure is impossible. The time evolution must take the two disjoint volumes of phase space associated with state $L$ and $R$ and evolve them to a single volume associated with the *reset* state. This many-to-one mapping in the phase space is precluded by the invertibility of the time evolution generated by the Hamiltonian.

The coarse graining of phase volumes escapes this difficulty and makes strong erasure possible. Using this conception, the Hamiltonian time evolution allows the phase volumes associated with each of the states $L$ and $R$ to evolve to disjointed volumes whose union, when coarse grained, represents a single state for both the system erased and the environment. This is shown in stylized form in Figure 5. The coarse-grained erasure state is the interleaved union of the two evolved states $L$ and $R$ on the right of the figure.

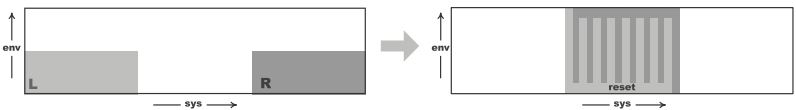

**Figure 5** A Coarse-grained Many-to-One Mapping.

The figure shows how the time evolution respects the conservation of phase volume required by the Liouville theorem. The phase volume contributed by the state $L$ to the reset state is the same in magnitude as state $L$'s initial phase volume; and state $R$ contributes a corresponding volume to the reset state. However, the *reset* state corresponds to a coarse-grained volume of phase space that is the union of the two component volumes. The coarse-grained volume is identified as all points in phase space whose system degrees of freedom conform with the *reset* state specification; and with the macroscopic variables that characterize the corresponding state of the environment after the erasure process. Typical erasure processes require a very slight heating of the environment, whose state is now characterized by a very slight increase in its temperature variable. Other environmental variables would include those that characterize the change of state of any machinery used to effect the erasure process.

---

6    Resist the temptation to identify the pre-erasure state with the union of phase space volumes $V_{ph,L}$ and $V_{ph,R}$, so that it becomes the thermalized state "$L + R$." They differ in their dynamic properties. If the phase point is momentarily in $L$ for the thermalized state, it may later be found in $R$, whereas this is impossible for pre-erasure state.

These last conclusions can be given simple quantitative expressions. The coarse-grained volume of the *reset* state must equal or exceed in magnitude the sum of the individual volumes that evolve from states *L* and *R*. For the phase volume of the totality—system plus environment—we have:

$$V_{ph,\text{reset}} \geq V_{ph,L} + V_{ph,R}. \quad (4)$$

Applying (1) to (4) we recover the minimum entropy cost of erasing each of the states *L* and *R* individually. That is,

$$\Delta S_L = S_{reset} - S_L = k \log(V_{ph,reset}/V_{ph,L}) \geq k \log((V_{ph,L} + V_{ph,R})/V_{ph,L}), \quad (5)$$

and similarly,

$$\Delta S_R \geq k \log((V_{ph,L} + V_{ph,R})/V_{ph,R}). \quad (6)$$

In anticipation of the information-theoretic ideas to be introduced in Section 6 below, we can take the case in which we are uncertain over which state is presented for erasure. We assign probability *p* to state *L*; and probability 1–*p* to state *R*. The probabilistically-weighted entropy cost of erasure is:

$$p\Delta S_L + (1-p)\Delta S_R \geq -k \left( p \log\left(\frac{V_{ph,L}}{V_{ph,L} + V_{ph,R}}\right) + (1-p) \log\left(\frac{V_{ph,R}}{V_{ph,L} + V_{ph,R}}\right) \right). \quad (7)$$

It follows that the information-theoretic entropy $S_{info}$, defined below in (16), is a minimum entropy cost of erasure only when the probability *p* is tuned to one specific value:

$$p = \left(\frac{V_{ph,L}}{V_{ph,L} + V_{ph,R}}\right). \quad (8)$$

The phase space and information theoretic analysis of Table 2 below considers the implications of this restriction in the value of *p* for the special case of $V_{ph,L} = V_{ph,R}$, when *p* = 1/2. This special case arises with a Szilard one-molecule gas, initially divided into equal cylinder volumes. Equations (5) and (6) entail an entropy cost of erasure of

$$\Delta S_L = \Delta S_R \geq k \log 2 \qquad Q_{env} \geq kT \log 2 \quad (9)$$

The environmental heating $Q_{env}$ follows when we assume that the environment is a heat bath at temperature *T* and that the reset state is one half the cylinder volume. These results are distinctive in specifying the entropy cost of erasure for each state presented individually.[7]

No procedure can realize these minima since, as we shall soon see, any such procedure must create further entropy to suppress fluctuations. However, if we neglect fluctuations, the following procedure, shown in Figure 6, realizes the minima (9) for the Szilard one-molecule gas:

1. Remove the partition.
2. Reversibly compress the gas to the reset state.

---

7   Norton (2013, 4445) noted that erasure does not require dissipation in so far as it only involves the relocation equal volumes of phase space. This note applies only to weak erasure.

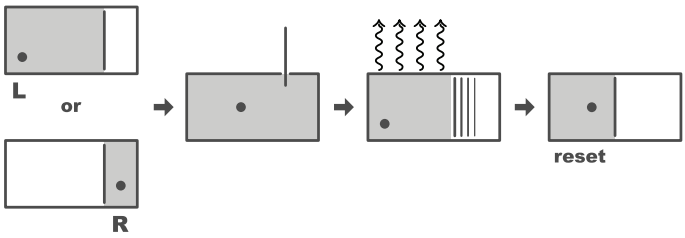

**Figure 6** Strong Erasure of a Szilard One-Molecule Gas System.

The irreversible Step 1 only creates entropy in the amount indicated by equality in relations (5), (6), and (9). This procedure conforms with *Szilard's condition*, since each step can be carried out independently of the physical state to be erased. It also conforms with *Bennett's condition*. The reversible compression of Step 2 passes the same quantity of heat to environment, independently of the physical state to be erased. If we assume, with (9), that the system reset state is one half of the cylinder volume, the heat passed is $Q_{env} = kT \log 2$.

For more general cases of erasure, without some further specification of the systems involved, we can only conjecture that *Szilard's condition* can be made to hold. *Bennett's condition* will hold since the coarse-grained state of the reset system and auxiliaries is the same for each state erased.

## 5. THE ENTROPY COST OF SUPPRESSING FLUCTUATIONS

The inequalities of (5), (6), and (9) specify the minimum entropy cost of erasure. It is easy to see that the dynamical character of thermal systems prevents these lower bounds from being realized or even approached. This follows from the fact that thermal processes only advance when they are entropically favored, without their completion being absolutely assured. The absolute completion of the process discussed in Section 4 is an aspiration that cannot be fully achieved. An ideal gas expands since the expanded state has greater entropy. But a very rare, random fluctuation can still spontaneously recompress it back to the lower entropy state. A particle that has fallen into a deep energy well can still escape if it momentarily and improbably gains enough energy from a heat bath. Completion at molecular scales is always only probabilistic.

The general result governing this behavior is given by (4): The evolution in time of the phase point in the total phase space is such that the probability of being in a given region of phase space is proportional to the phase volume of the region, as shown in Figure 7.

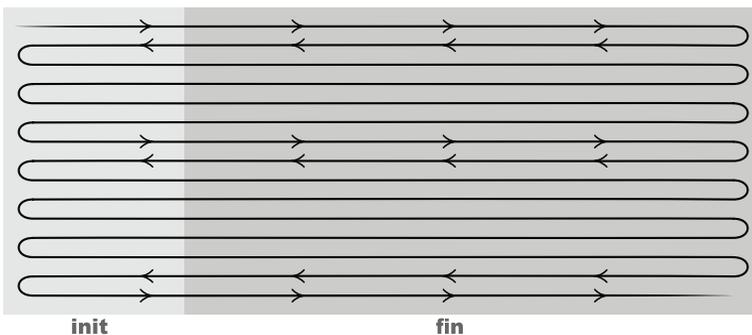

**Figure 7** Occupation Times are Proportional to Phase Volume.

Thus, if a process is to advance from some initial state "*init*" to a final state "*fin*," the phase volume of the final state must be significantly larger than that of initial state. Only then can completion of the process be assured, and even then only probabilistically. There will always be some small probability that its migration takes the phase point back to the initial state. This expansion of the phase volume of the final state corresponds to the creation of entropy. The greater the entropy created, the more dissipative is the process, but the more probable is its completion.

The connection between phase volume and probability (4), combined with (1) above, yields what Einstein called "Boltzmann's principle" or "$S = k \log W$." It connects the entropy $S$ of a system with its probability, $P$. Applied to the above process, Boltzmann's principle asserts

$$\Delta S = S_{fin} - S_{init} = k \log (P_{fin}/P_{init}) \quad \text{or} \quad P_{fin}/P_{init} = \exp(\Delta S/k). \quad (10)$$

This principle forces us to trade-off entropy creation against the probability of completion in processes on molecular scales, where entropies of a few $k$ are significant. Take, for example, a process driven by an entropy increase:

$$\Delta S = S_{fin} - S_{init} = k \log 2,$$

such as is common in Szilard's problem. If this is the only entropy increase in the erasure process, then its completion is compromised. That is, we have from (10) that

$$P_{fin}/P_{init} = \exp(\Delta S/k) = \exp(\log 2) = 2.$$

At any moment, the probability that erasure has been successfully completed is only twice the probability that the system has reverted by a fluctuation to the original, unerased state.

We need processes that are substantially more dissipative if we are to secure probabilistic completion of processes on molecular scales. That requires a coarse-grained reset state of substantially larger phase volume than the sum of $V_{ph,L}$ and $V_{ph,R}$, as shown in the stylized Figure 8.

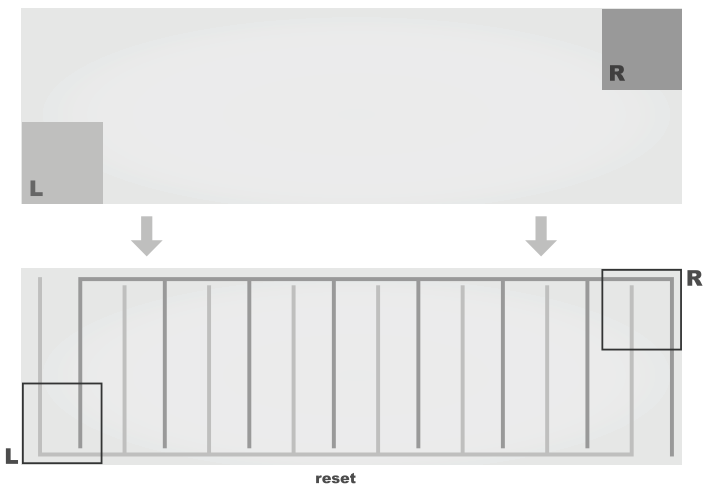

**Figure 8** Expanded Reset State is Probabilistically Favored.

In erasure, the phase volumes of states $L$ and $R$ are conserved, but their volumes are stretched into thin tendrils spread over the whole reset space. The coarse-grained reset state is the entirety of the rectangle in Figure 8 within which the evolved states $L$ and $R$ are found. As a phase point initially in $L$ explores the tendrils, it spends much more time in the large phase volume associated with the reset state than in the smaller volume of the initial, unerased state $L$ (and similarly for phase points initially in $R$).

A modest probability ratio of only $P_{fin}/P_{init} = 20$ requires a twenty-fold increase in phase volume and an entropy creation of $k \log 20 = 3k$. Since the ratio of probabilities increases exponentially with entropy difference, the ratio rapidly grows large with modest increases in entropy creation and ceases to be a problem, outside the realm of molecular-scale processes.

These probabilistic disturbance to processes may seem abstruse. They are, however, familiar effects in thermal systems and are otherwise known as thermal fluctuations or, in electrical engineering, noise or static. They cannot be idealized away since they are intrinsic to the dynamical character of thermal properties. Two systems are in thermal equilibrium only when they are exchanging energy dynamically. Fluctuations—momentary imbalances—are an ineliminable feature of those exchanges. Norton (2011, 2013, 2017) has computed many examples of fluctuations and the entropic cost of their suppression.

## 6. THE INFORMATION-THEORETIC ANALYSIS

While the phase space analysis above gives a compact and serviceable analysis of the entropic costs of erasure, by far the more common analysis uses information-theoretic ideas.[8] That is, if we have a system that may be in either of two mutually exclusive states, $L$ or $R$, but we know not which, an additional thermodynamic entropy (15) below is assigned to the system as a result of our lack of information.[9] Erasure eliminates this lack of information and the thermal cost of erasure is determined from the ensuing decrease in the system's thermodynamic entropy.

### 6.1 INTRODUCING INFORMATION ENTROPY

States $L$ and $R$ occupy disjoint phase spaces $\Gamma_L$ and $\Gamma_R$, where these spaces comprise only the degrees of freedom of the system $L$ and $R$, excluding the degrees of freedom of the environment. Their union "$L + R$" occupies phase space $\Gamma_{L+R} = \Gamma_L \cup \Gamma_R$. Their phase points are canonically distributed as:

$$\rho_L(x) = \exp(-E(x)/kT) / Z_L \text{ for } x \in \Gamma_L$$
$$\rho_R(x) = \exp(-E(x)/kT) / Z_R \text{ for } x \in \Gamma_R$$
$$\rho_{L+R}(x) = \exp(-E(x)/kT) / Z_{L+R} \text{ for } x \in \Gamma_{L+R} \quad (11)$$

where $E(x)$ is the energy at phase point $x$ and the normalizing partition functions are

$$Z_L = \int_{\Gamma_L} \exp\left(-\frac{E(x)}{kT}\right) dx \quad Z_R = \int_{\Gamma_R} \exp\left(-\frac{E(x)}{kT}\right) dx \quad Z_{L+R} \int_{\Gamma_{L+R}} \exp\left(-\frac{E(x)}{kT}\right) dx. \quad (12)$$

Prior to erasure, the system is in one of states $L$ or $R$. If, for example, the system is a Szilard one-molecule gas, the molecule is assuredly trapped by a partition on either the left or right side of the chamber, we know not which. This compounded state is represented by a weighted sum of the distributions:

---

8   For historical surveys of the earlier years, see Earman and Norton (1998, 1999) and Leff and Rex (2003).

9   This added probability is epistemic and does not conform with the dynamic conception of probability of condition (4) above.

$$\rho_{comp}(x) = p\rho_L(x) + (1-p)\rho_R(x), \quad (13)$$

where $0 < p < 1$ is a weight that may be an epistemic probability or a reflection of the rate of occurrence of the states.

The Gibbs entropy formula:

$$S(\rho) = -k \int_\Gamma \rho \log \rho \, dx, \quad (14)$$

is applied to (13) to recover the entropy of the compound state:

$$\begin{aligned} S_{comp} &= -k \int_{\Gamma_{L+R}} \rho_{comp} \log \rho_{comp} \, dx \\ &= -pk \int_{\Gamma_L} \rho_L \log \rho_L \, dx - (1-p)k \int_{\Gamma_R} \rho_R \log \rho_R \, dx - k(p \log p + (1-p)\log(1-p)) \\ &= pS_L + (1-p)S_R - k(p \log p - (1-p)\log(1-p)). \end{aligned} \quad (15)$$

The third term in (15), an "information entropy" term, is reminiscent of Shannon's information theory:

$$S_{info} = -k\big(p \log p + (1-p)\log(1-p)\big). \quad (16)$$

The simplest case arises when entropies of the states *L*, *R,* and *reset* are equal, so that

$$S_L = S_R = S_{reset}$$

This is, for example, the case of a Szilard one-molecule gas divided into equal volumes and then erased to *L*. In this case, the entropy change in the system upon erasure is

$$\Delta S_{sys} = S_L - S_{comp} = -S_{info} < 0. \quad (17)$$

Since total entropy $S_{tot}$ cannot decrease, it follows that the entropy of the environment increases by at least $S_{info}$. When the environment is represented by a heat bath at temperature *T*, this entropy increase corresponds to an environmental heat gain $Q_{env}$ of at least $TS_{info}$. In sum, the dissipation associated with the erasure of the compound state is

$$\Delta S_{tot} \geq 0 \quad \Delta S_{env} \geq S_{info} \quad Q_{env} \geq TS_{info} = -kT(p \log p + (1-p)\log(1-p)), \quad (18)$$

for $p = 1/2$, $S_{info}$ takes its maximum value of $k \log 2$ and $Q_{env} \geq kT \log 2$.

## 6.2 ITS PROBLEMS

There are significant problems with these results. The most significant is that the lower bounds of (18) are unattainable. The information-theoretic analysis has neglected the dissipation arising from the need to suppress fluctuations.

If we set aside fluctuations and consider only the dissipation associated with many-to-one mappings, these results are still inconsistent with the phase space analysis of erasure. Perhaps the most striking difference is that erasure in this information-theoretic analysis is not dissipative in the familiar sense of creating thermodynamic entropy. Rather, dissipation arises only in the sense that entropy is moved in a thermodynamically reversible process from the system to the environment, which results in a heating of the environment.

While this may seem unremarkable, it renders the information-theoretic approach incompatible with a simple formulation of what is called the "The Thermodynamics of Computing." That simple formulation depends on an equation: Logically reversible computations are implemented by thermodynamically reversible processes; and, logically irreversible computations, such as erasure, are implemented by thermodynamically irreversible processes. While Bennett's (1982) is a standard presentation, the simple formulation is not endorsed by him. See Bennett (2003, 502).[10]

The information-theoretic conception of erasure is one of strong erasure in so far as it satisfies *Bennett's condition* in passing the same quantities of heat (18) to environment, independently of the state erased. However, one reading of (18) is weak erasure. In it, these quantities are averages over many cases, so that differential heating of the environment may leave a trace of the state erased. Below, the information-theoretic analysis is compared with the phase space analysis for both weak and strong conceptions in Tables 1 and 2. There are mismatches in both cases.

| PHASE SPACE ANALYSIS | INFORMATION-THEORETIC ANALYSIS |
|---|---|
| Minimum total entropy change is zero. | Minimum total entropy change is zero. |
| Minimum entropy change for system and environment individually is zero. | Minimum entropy change for system is $-S_{info}$ and for the environment is $S_{info}$. |
| Minimum heat passed to the environment is zero. | Minimum heat passed to the environment is $T S_{info} = -kT(p \log p + (1-p) \log(1-p))$ and varies from 0 to $kT \log 2$ depending on the value of $p$. |
| Results are independent of parameter $p$ | Results depend on parameter $p$ |

Table 1 Comparison for Weak Erasure.

| PHASE SPACE ANALYSIS | INFORMATION-THEORETIC ANALYSIS |
|---|---|
| Minimum total entropy change is $k \log 2$. | Minimum total entropy change is zero. |
| Minimum total entropy change $k \log 2$ applies to erasure of each state $L$ and $R$ individually. | Minimum entropy change for the system is $-S_{info}$ and for the environment is $S_{info}$. |
| Minimum heat passed to the environment is $kT \log 2$. | Minimum heat passed to the environment is $T S_{info} = -kT(p \log p + (1-p) \log(1-p))$ and varies from 0 to $kT \log 2$ depending on the value of $p$. |
| Results are independent of parameter $p$. | Results depend on parameter $p$. |

Table 2 Comparison for Strong Erasure.

To adjudicate the difference, we ask after the commonly discussed but fictional sorts of procedures applied to the Szilard one-molecule gas. Are there any that can realize these minima in the quantity observable through its heating effect, that is, through the heats passed to the environment? There is, as far as I know, no procedure that realizes the smaller minima, (18), when $p$ differs from 1/2. To get a sense of the difficulty of finding a such a procedure, consider a simple candidate for the case of $p > 1/2$, shown in Figure 9:

1. Reversibly move the partition rightwards from its position at half the volume to the larger $p$th fraction.
2. Remove the partition.
3. Reversibly compress the gas to the reset state of half the cylinder volume.

---

10  Erasure of data is thermodynamically reversible or irreversible, Bennett (2003) asserts, according to whether the data is "unknown" or "known," respectively. In the first case of unknown data, thermodynamic reversibility is possible, since erasure is conceived as the conveyance of entropy—presumably the information entropy—from the system to the environment. In the second case, since there is no information entropy, erasure is conceived as thermodynamically irreversible, although it can be made reversible by the strategy of recording a trace of the data elsewhere.

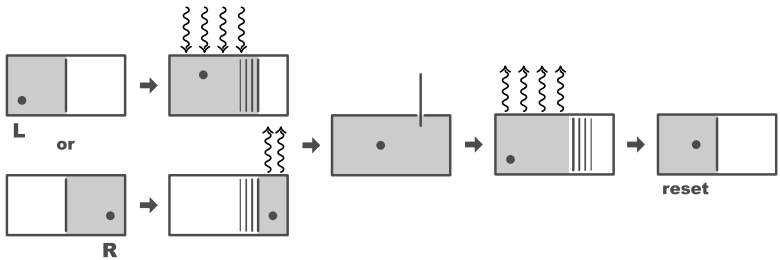

**Figure 9** Erasure of Szilard One-Molecule Gas System for Parameter $p$.

This procedure is not a candidate for strong erasure, but only for weak erasure, since it violates *Bennett's condition*. Different quantities of heat are passed to the environment according to which state is erased. It passes a net heat $-kT \log p$ if $L$ is erased and a net heat of $-kT \log(1-p)$ if $R$ is erased.[11] If these two quantities are weighted with factors $p$ and $(1-p)$ and summed, we recover $TS_{info}$, which is the minimum heat transfer to the environment of (18). The more serious problem is that Step 1. violates *Szilard's condition*. Different apparatus are needed according to whether the gas is in $L$ and Step 1. expands the gas, or the gas is in $R$ and Step 1. compresses it. We may conjecture that a more elaborate procedure can conform with *Szilard's condition* and perhaps even *Bennett's condition*. If, however, the phase space analysis is correct for strong erasure, no such elaboration can succeed for strong erasure and the minimum environmental heating is just $kT \log 2$.

If this last problem cannot be resolved, the entire rationale of the information-theoretic approach is undercut. The rationale is that erasure is thermodynamically costly because of our lack of information. The worse informed we are, supposedly the greater the cost. The extent of our lack of information is measured by the information entropy term (16), which also fixes the amount of thermal dissipation. The worst case is $p = 1/2$, in which we are maximally unsure of which state is to be erased and (16) takes its maximum value. As $p$ approaches 0 or 1, we become better informed as to which of $L$ or $R$ is to be erased. Now better informed, we should be able to erase with less dissipation, since the information entropy term (16) decreases to zero as $p$ approaches these limits. However there seems to be no way to realize this lesser dissipation in strong erasure for more favorable values of $p$.

If the phase space analysis is correct, the parameter $p$ has no place in the analysis at all, either as an epistemic probability or as a frequency of occurrence of states. The amount of dissipation derives only from the requirement that a single procedure must work equally on either of the two states presented for erasure in the one case at hand.

## 7. INFORMATION ENTROPY IS NOT GIBBS THERMODYNAMIC ENTROPY: THE FALLACY

The most serious problem facing the information-theoretic analysis is that the above introduction of the information-theoretic entropy term (16) is fallacious. The application of the Gibbs entropy formula (14) to the compound state (13) in the computation (15) is a misapplication of the Gibbs formalism. The full Gibbs entropy formula contains an additive constant whose evaluation leads to the elimination of the information entropy term in (15).

---

11   Step 1. passes heat $-kT \log 2p$ to the environment, if $L$ is erased; and $-kT \log 2(1-p)$ if $R$ is erased. The net heats transferred are recovered by adding the heat $kT \log 2$ passed to the environment in Step 3.

## 7.1 DERIVING GIBBS ENTROPY

The Gibbs formalism, as developed in Gibbs (1914) and Einstein (1903), applies specifically to a canonically distributed system, such as in (11). It seeks to identify quantities that play the role of temperature, entropy, and the like in the statistical analysis by matching them with analogous terms in the thermodynamic analysis. A correlate of the Clausius entropy should match in two properties:

- Changes in expectation in this quantity correspond in reversible processes to the systems' incremental gain in heat, divided by temperature; and

- Irreversible processes, driven by imbalanced generalized forces, correspond to those that increase the totality of this quantity.

Following the summary given in Norton (2005, §2.2), the change in the system's mean energy $\bar{E}$ is determined under slow changes of the temperature $T(t)$ and the Hamiltonian $E(x, \lambda(t))$, where the changes are tracked by a path parameter $t$ that affects the Hamiltonian through a parameter $\lambda(t)$. The rate of change of the mean energy is given by:

$$\frac{d\bar{E}}{dt} = \frac{d}{dt}\int_\Gamma E(x,\lambda)\rho(x,t)dx = \int_\Gamma \frac{dE(x,\lambda)}{dt}\rho(x,t)\,dx + \int_\Gamma E(x,\lambda)\frac{d\rho(x,t)}{dt}\,dx.$$

The first term in the sum is identified as the rate at which work is done on the system. Comparing this expression with the thermodynamic equality change in internal energy = work done on system + heat gained by system,

the second term is identified as the mean rate at which the system gains heat $Q$:

$$\frac{dQ}{dt} = \int_\Gamma E(x,\lambda)\frac{d\rho(x,t)}{dt}\,dx.$$

Since this is a reversible process, we can use Clausius' definition of entropy, $dS = dQ_{rev}/T$, to introduce the thermodynamic entropy in terms of the mean heat gain $Q_{rev}$ as:

$$\frac{dS}{dt} = \frac{1}{T}\frac{dQ_{rev}}{dt} = \frac{1}{T}\int_\Gamma E(x,\lambda)\frac{d\rho(x,t)}{dt}\,dx = \frac{d}{dt}\left(\frac{\bar{E}}{T} + k\log Z(t)\right).$$

The last equality is recovered only after considerable manipulation. Integrating, we recover the expression for the canonical entropy:

$$S = \frac{\bar{E}}{T} + k\log Z + constant,$$

where the constant is independent of the variables altered in the reversible process with path parameter $t$.

This canonical expression is the one derived by Gibbs (1914, 44) and Einstein (1903, 182) and in subsequent developments of their work, such as Tolman (1927, 302–303). Recalling that the mean energy $\bar{E}$ and the partition function $Z$ derive from the canonical distribution (11), this canonical entropy is equivalent to

$$S = -k\int_\Gamma \rho\log\rho\,dx + constant, \qquad (19)$$

Expressions like these appear in Gibbs' analysis (e.g., 1914, 136) and in the Ehrenfests' (1911, 51, 54, 61) comparison of Boltzmann's and Gibbs' developments. The unqualified identification of this expression as the "Gibbs entropy" comes much later in the history and may even be as late as Jaynes (1965).

## 7.2 GIBBS ENTROPY OF A COMPOUND STATE

The derivation of the Gibbs entropy formula (19) assumes throughout that the probability distribution is canonical, that is, has the form $\exp(-E(x)/kT)/Z$. In general, a compound probability distribution such as (13) does not have this form. It will only do so when the parameter $p$ is adapted to the states $L$ and $R$ by

$$p = Z_L/(Z_L + Z_R) \text{ and } (1-p) = Z_R/(Z_L + Z_R), \tag{20}$$

for then,

$$\rho_{comp}(x) = p\rho_L(x) + (1-p)\rho_R(x)$$

$$= \frac{Z_L}{Z_L + Z_R} \cdot \frac{\exp\left(-\frac{E}{kT}\right)}{Z_L}\bigg|_{\Gamma_L} + \frac{Z_R}{Z_L + Z_R} \cdot \frac{\exp\left(-\frac{E}{kT}\right)}{Z_R}\bigg|_{\Gamma_R} = \frac{\exp\left(-\frac{E}{kT}\right)}{Z_{L+R}}.$$

With $p$ adapted to the states $L$ and $R$, the Gibbs entropy formula (19) can be applied to a compound state (13) and, using computations analogous to (15), gives:[12]

$$S_{comp} = -k \int_{\Gamma_{L+R}} \rho_{comp} \log \rho_{comp}\, dx + \text{constant}$$
$$= pS_L + (1-p)S_R - k\big(p \log p + (1-p)\log(1-p)\big) + \text{constant}. \tag{21}$$

## 7.3 COMPATIBILITY OF ZERO STATES FOR ENTROPIES OF SIMPLE AND COMPOUND SYSTEMS

The presence of the constant in the canonical entropy and Gibbs entropy formulae is not generally noted. In familiar, simple states, such as a gas confined to a chamber, it is easily seen that it plays no role in the physics. It can be set to zero, which is the setting assumed for the following.

Matters become more delicate when we compare the entropies of different types of systems, such as a simple state and a compound state. While, overall, we can always set an arbitrary zero point for entropies, we must ensure that the entropies of simple and compound states are set with compatible zero points. Otherwise, we risk spurious terms confounding the comparison of the entropies of simple and compound states. To preclude this error, we continue the Einstein-Gibbs method of matching statistical quantities with analogous quantities in thermodynamics.

We can arrive at a compatible zero point for the entropies of simple and compound systems if we consider a process that connects them. It is the removal of the partition in the case of a Szilard one-molecule gas (and its analog for more general systems). That process precludes a zero value for the constant in (21) for compound states. For if we set the constant to zero, then the entropy of the compound system (13) is equal to the entropy of the thermalized system, that is, of the system "$L + R$" of (11) prior to insertion of the partition:

$$S_{comp} = S_{L+R}.$$

---

12   This last consideration does not preclude application of the Gibbs entropy formula to other distributions. However, if the entropy recovered is to relate to the Clausius entropy $dS = dQ_{rev}/T$, then a new justification beyond those of Gibbs and Einstein is needed. That such a justification is possible is suggested by the fact that a process that alters the entropies of states $L$ and $R$ in (21) by $\Delta S_L$ and $\Delta S_R$ leads to a new entropy $S_{comp} = p(S_L + \Delta S_L) + (1-p)(S_R + \Delta S_R)$, which still has the form (23) below, even though $p$ may not be adapted to the new states $L$ and $R$ by (20).

This follows immediately from the Gibbs entropy formula, since the distribution (13) for the compound system adapted to the states by (20) is the same as that for the thermalized system in (11), so that $\rho_{comp}(x) = \rho_{L+R}(x)$.

Consider, thermodynamically, the process that ensues after removal of the partition in Szilard's one-molecule gas. We momentarily have a one-molecule gas confined to one or the other side of the chamber. It will expand irreversibly to fill the chamber. Such expansion is an elementary example of an irreversible process in thermodynamics. If we have set the constant in (21) to zero for the compound state, then the momentarily compressed state and the thermalized state have the same entropy. In the absence of an entropic driving force, the two states are at equilibrium and we should not expect that one will evolve into the other.

This equality of entropies violates the fundamental notion of thermal dynamics, that changes are driven by entropy differences. If we are to preserve this notion, we must choose a different value for the constant in (21). The natural choice is

$$\text{constant} = +k\Big(p \log p + (1-p) \log (1-p)\Big). \tag{22}$$

With this choice, the entropy assigned to the compound system is merely

$$S_{comp} = p\, S_L + (1-p)\, S_R. \tag{23}$$

It is the natural value for the entropy. For, if we treat entropy like other extensive magnitudes in thermodynamics such as internal energy, we would expect the compounded value simply to be the weighted sum of the component magnitudes. The entropy of the thermalized state becomes

$$S_{therm} = pS_L + (1-p)S_R - k(p \log p_L + (1-p)\log(1-p)) > S_{comp}.$$

Thus, the entropy of the thermalized state now exceeds that of the compound state by $-k(p \log p + (1-p) \log(1-p))$ and this entropy difference drives the irreversible process that takes the compound state to the thermalized state.

## 7.4 THE COMPOUND STATE IS A FLAWED CONCEPTION

These last considerations render unsustainable the information entropy term (16) in the expression (15) for the entropy of a compound state. However, they only make it "natural" to choose the specific value (22) for the constant that leads to the weighted sum of entropies (23). A simpler consideration indicates that (23) is the uniquely correct expression. It arises at the starting point of the information-theoretic analysis. Our goal at the outset is to find a way to represent our uncertainty over which of states $L$ or $R$ are present, using the parameter $p$.

If our concern is the entropy or energy or any other extensive magnitude among the states present, there is no other choice beyond a $p$ weighted sum of the form (23). If $p$ is read as a frequency of occurrence of the various states, then the $p$ weighted sum of (23) simply is the average value of the entropy over many cases. If $p$ is an epistemic probability, then (23) is the expectation value of the entropy. This is where the analysis should have started.

To start with the compounded distribution (13) as representing our uncertainty is an invitation for fallacy and confusion, for the compounding merges probabilities of different types. The probabilities of the canonical distributions $\rho_L(x)$ and $\rho_R(x)$ of (11) are dynamical and track the changes over time of the state of each system. They are the bearers of thermodynamic properties. The parameter $p$, introduced as a probability measure over the two canonical distributions $\rho_L(x)$ and $\rho_R(x)$, is static. It is set at the outset externally by us and should not be presumed automatically to bear thermodynamic properties. Once the two are merged, we have a dangerous, blended measure that is neither a purely epistemic probability nor a purely thermodynamic probability. That does not preclude further computations with this hybrid structure. But it does mean that all such computations must proceed with the most extreme caution if a fallacy is to be avoided. The greatest danger is that thermodynamic properties are attributed incorrectly to the static probability *p*. The analysis of this section shows that the literature has not proceeded with the requisite caution and has committed precisely this fallacy.

There is, to my mind, something already dubious in the introduction of the parameter *p*. It is an additional term not present in the thermodynamics of the systems to be erased. Our circumstance is merely that we do not know which state is present. The phase space analysis shows that we can have a simple and serviceable analysis of erasure on that basis alone without any appearance of a "*p*." We may hope that the introduction of the parameter *p* would be a benign detour whose influence can be eliminated. The accretion of problems for the information-theoretic analysis indicates otherwise.

## 8. CONCLUSION

On molecular scales, the dominant source of dissipation lies in the entropy creation needed to suppress thermal fluctuations and assure probabilistic completion of all processes of any type. In the case of erasure, there is a second, lesser source of dissipation that derives from the character of erasure itself as a many-to-one mapping. A major concern of this paper has been to determine the magnitude of this dissipation and to find its origin.

We have seen two competing analyses. The information-theoretic analysis locates this origin in the pre-erasure state itself. It attributes an additional thermodynamic entropy to this state that arises merely from our ignorance over which state is present for erasure. Dissipation consists merely in the passage of this extra entropy to the environment in what may otherwise be a thermodynamically reversible process.

The analysis fails in several ways. It indicates minima of dissipation in erasure that varies according to the extent of our ignorance, even though most of the minima appear unachievable in the case of strong erasure. More seriously, the attribution of this additional entropy is derived fallaciously from a misapplication of the Gibbs formalism that leads to a mistaken identification of information entropy and thermodynamic entropy.

The phase space analysis does not assign any increase in the entropy of the pre-erasure states from our uncertainty over which is present. Instead, the entropy cost of erasure arises from the core restriction that a single procedure must be employed in erasure, independently of the states presented for erasure what we may know of them.

## ACKNOWLEDGEMENTS

This paper developed in the course of extensive discussions with Jacob Barandes and Wayne Myrvold, whose stimulation is gratefully acknowledged but their agreement is not presumed.

## COMPETING INTERESTS

## AUTHOR AFFILIATIONS
**John D. Norton**  orcid.org/0000-0003-0936-5308

Department of History and Philosophy of Science, University of Pittsburgh, Pittsburgh, PA, USA